RESEARCH ARTICLE

# Drug-induced activation of integrin alpha IIb beta 3 leads to minor localized structural changes

Una Janke[1,2], Martin Kulke[1], Ina Buchholz[1,2], Norman Geist[1], Walter Langel[1], Mihaela Delcea[1,2,3] *

1 Institute of Biochemistry, University of Greifswald, Felix-Hausdorff-Straße 4, Greifswald, Germany, 2 ZIK HIKE- Zentrum für Innovationskompetenz "Humorale Immunreaktionen bei kardiovaskulären Erkrankungen", University of Greifswald, Fleischmannstraße 42, Greifswald, Germany, 3 DZHK (German Centre for Cardiovascular Research), partner site Greifswald, Germany

* delceam@uni-greifswald.de

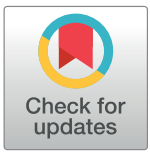

Data Availability Statement: The pdb file of the integrin structures used for MDS is available at the following link: https://www.researchgate.net/publication/332061683_Integrin_alphaIIb_beta3.

Funding: This work was financially supported by the European Research Council (ERC) Starting Grant 'PredicTOOL' (637877) to M.D.

Competing interests: The authors have declared that no competing interests exist.

## Abstract

Integrins are transmembrane proteins involved in hemostasis, wound healing, immunity and cancer. In response to intracellular signals and ligand binding, integrins adopt different conformations: the bent (resting) form; the intermediate extended form; and the ligand-occupied active form. An integrin undergoing such conformational dynamics is the heterodimeric platelet receptor αIIbβ3. Although the dramatic rearrangement of the overall structure of αIIbβ3 during the activation process is potentially related to changes in the protein secondary structure, this has not been investigated so far in a membrane environment. Here we examine the $Mn^{2+}$- and drug-induced activation of αIIbβ3 and the impact on the structure of this protein reconstituted into liposomes. By quartz crystal microbalance with dissipation monitoring and activation assays we show that $Mn^{2+}$ induces binding of the conformation-specific antibody PAC-1, which only recognizes the extended, active integrin. Circular dichroism spectroscopy reveals, however, that $Mn^{2+}$-treatment does not induce major secondary structural changes of αIIbβ3. Similarly, we found that treatment with clinically relevant drugs (e.g. quinine) led to the activation of αIIbβ3 without significant changes in protein secondary structure. Molecular dynamics simulation studies revealed minor local changes in the beta-sheet probability of several extracellular domains of the integrin. Our experimental setup represents a new approach to study transmembrane proteins, especially integrins, in a membrane environment and opens a new way for testing drug binding to integrins under clinically relevant conditions.

## Introduction

The heterodimeric platelet receptor integrin αIIbβ3 mediates cell adhesion and plays a critical role in hemostasis and clot formation [1, 2]. Therefore, regulating the activity of αIIbβ3 is essential for platelet stimulation and prevention of their uncontrolled aggregation [3, 4]. The expression of αIIbβ3 is restricted to megakaryocytes, where the two subunits are assembled in the endoplasmic reticulum. After post-translational processing in the Golgi apparatus, the 235 kDa protein, comprised of 1827 amino acids, translocates to the platelet surface and is expressed with approximately 80,000 copies per platelet [5].

Integrin αIIbβ3 is a bidirectional receptor that undergoes *outside-in* and *inside-out* signaling and is present in at least three different conformations as demonstrated by cryo-electron microscopy (EM), negatively stained EM or by nuclear magnetic resonance [6–8]: i) the bent (resting) low affinity state; ii) the intermediate extended state (opening); and iii) the ligand-occupied high affinity active form [9].

Divalent ions have been widely shown to be essential for integrin function, stabilization of subunit interaction, regulation of ligand binding and consequently for the activation of the protein. The addition of EDTA removes divalent cations (i.e. $Ca^{2+}$, $Mg^{2+}$) from their binding sites and leads to the inhibition of integrin-ligand binding [10]. By contrast, the non-physiological stimulation by manganese ions ($Mn^{2+}$) shifts integrin into its high affinity conformation by potential opening of the hinge angle at the hybrid domain and alters the cation coordination in the β3 A-domain by binding to the metal ion-dependent adhesion site (MIDAS) [2, 10, 11], as illustrated in Fig 1.

However, the physiological activation of αIIbβ3 in platelets is mediated by talin-1 that links the integrin cytoplasmic domain to the actin cytoskeleton and initiates unclasping between the two cytoplasmic tails of αIIb- and β3-subunits. Talin-1 triggers conformational changes in the extracellular domain, which is associated with the *inside-out* signaling [11], such as the translocation of helix-7 within β3 A-domain and the repositioning of the plexin-semaphorin-integrin (PSI) and the hybrid domain [2, 8, 11]. After opening, integrin is able to bind e.g. fibrinogen *via* the RGD (arginine-glycine-aspartate) binding pocket, and this activates intracellular signaling pathways and leads to platelet aggregation (*outside-in* signaling) [12]. Besides these major changes in the tertiary structure of αIIbβ3, the transition between the position of the different domains during activation could lead to alterations in secondary structure. However, this was not studied so far in a membrane environment.

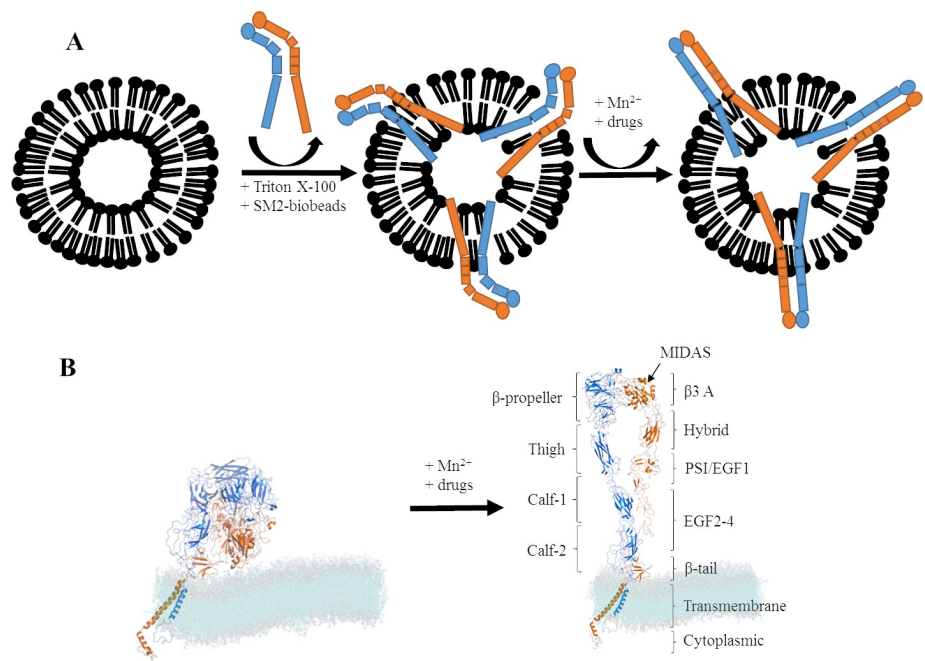

**Fig 1. Schematic illustration of the proteoliposomes and structure of integrin αIIbβ3.** A) Schematic illustration of the proteoliposomes as obtained after the reconstitution procedure before adsorption on $SiO_2$ surface. αIIbβ3 (αIIb-subunit in blue and β3-subunit in orange) is reconstituted into liposomes and treated with Triton X-100 as well as biobeads and activated by manganese ions ($Mn^{2+}$) or drugs. B) Structure of αIIbβ3 in bent (left) and open/active (right) conformation in a DMPG:DMPC (1:20) lipid membrane (cyan). The integrin model combines the αIIbβ3 transmembrane domain (PDB-code 2k9j) and ectodomain (PDB-code 3fcs), missing residues were added as random coils. The VMD 1.9. and PyMOL 2.1. softwarepackages were used to create this figure.

The conformation-specific antibody PAC-1 (IgM) binds only to the activated integrin αIIbβ3. The binding takes place at the fibrinogen binding site in the head domain formed by α and β subunits [13, 14]. Bhoria et al. used this antibody PAC-1 to show that platelets from patients affected by immune thrombocytopenia (ITP) exist in an activated state [15, 16]. Besides the platelet glycoprotein Ibα, integrin αIIbβ3 is the major antigen targeted by autoantibodies in ITP. Additionally, in secondary ITP, several drugs are potential candidates for inducing conformational changes in αIIbβ3. Examples are the anti-malaria drug quinine, or αIIbβ3 inhibitors such as eptifibatide, tirofiban and abciximab [17]. Similarly, unfractionated heparin (UFH), which is widely used as an anticoagulant in clinics, causes activation of αIIbβ3, platelet aggregation and increased affinity to its physiological ligand fibrinogen [18–20]. UFH, low-molecular weight heparins and the synthetic pentasaccharide fondaparinux induce *outside-in* signaling and consequently lead to platelet activation [19, 20]. Heparin binding is blocked by αIIbβ3 antagonists [21, 22]. This indicates that αIIbβ3 is activated by drugs, and this interaction induces structural changes, activation or even expression of cryptic epitopes (i.e. hidden binding sites) of the integrin.

Here, we study in a membrane environment the activation of αIIbβ3 reconstituted into liposomes, upon treatment with $Mn^{2+}$, quinine and heparins and investigate the related changes in the protein structure using various biophysical methods and molecular dynamics simulations.

## Results

αIIbβ3 was reconstituted into a membrane environment (i.e. liposomes) and treated with $Mn^{2+}$ or clinically relevant drugs (UFH, fondaparinux and quinine), as shown schematically in Fig 1A. The activation (opening) of integrin (Fig 1B) was determined by the binding of the conformation-specific antibody PAC-1 that is only interacting with the extended, active form of αIIbβ3.

### Validation of αIIbβ3 reconstitution into liposomes

The successful protein reconstitution into liposomes was validated by several techniques. Fig 2A presents TEM images of DMPG:DMPC vesicles with reconstituted protein (proteoliposomes). The inset shows spherical proteoliposomes with visible globular heads and stalk domains of the ectodomain of αIIbβ3 at the rim as indicated by the black arrow. By dynamic light scattering (DLS) it was demonstrated that the diameter of the proteoliposomes (255 ± 16.6 nm) was significantly larger than that of bare liposomes (161 ± 1.3 nm) (Fig 2B). The presence of the αIIb- and/or β3-subunits in the liposomes was verified by flow cytometry (Fig 2C) and SDS-PAGE (Fig 2D). Bare fluorescently labelled PE CF-liposomes show low unspecific staining by anti-CD41 antibody (anti-αIIb-subunit) and anti-CD61 antibody (anti-β3-subunit), whereas over 80% of the PE CF-proteoliposomes are CD41 and CD61 positive (Fig 2C). Moreover, both subunits migrated in a denaturing SDS-PAGE as two visible bands at 105 kDa (red arrow, αIIb-subunit) and 90 kDa (blue arrow, β3-subunit) in the proteoliposome sample, but not in the bare liposome fraction (Fig 2D).

### Monitoring the $Mn^{2+}$-induced activation of αIIbβ3 in proteoliposomes by different methods

We further addressed the activation state of αIIbβ3 in the liposomes by an activation assay with the conformation-specific antibody PAC-1. Fig 3A shows a substantially greater binding of PE CF-fluorescently labelled proteoliposomes to PAC-1 coated on a microtiter plate after

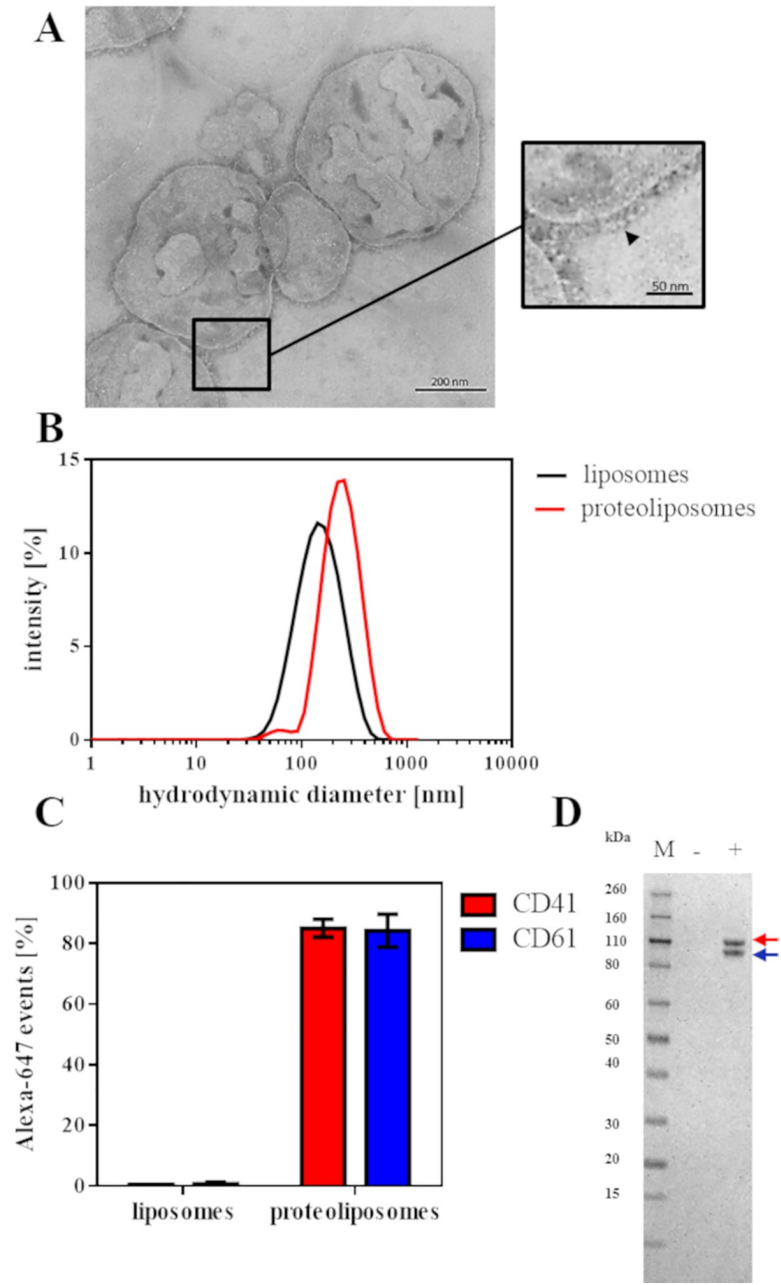

**Fig 2. Validation of αIIbβ3 reconstitution into liposomes.** A) TEM images of proteoliposomes. The inset shows a close-up view of αIIbβ3 (indicated by arrow) incorporated in a membrane environment. B) DLS data showing the hydrodynamic diameter of liposomes (black) and proteoliposomes (red) of three independent experiments measured in liposome buffer at 37°C. C) Statistical analysis of FACS-plots with liposomes and proteoliposomes from three independent measurements. Percentages of the mean ± standard error of the mean (SEM) of anti-CD41 (red) and anti-CD61 binding (blue) on PE CF- liposomes were plotted. D) Reductive SDS-PAGE of liposomes (-) and proteoliposomes (+), protein molecular weight standard (M) is shown on the left. The bands corresponding to the αIIb- (red) and β3-subunit (blue) are indicated by arrows.

incubation with $Mn^{2+}$ (red) compared to EDTA-treated proteoliposomes (blue). However, also without any additives, binding of PAC-1 to proteoliposomes was increased. By contrast, the flow cytometry measurements with PE CF- proteoliposomes, standard buffer conditions

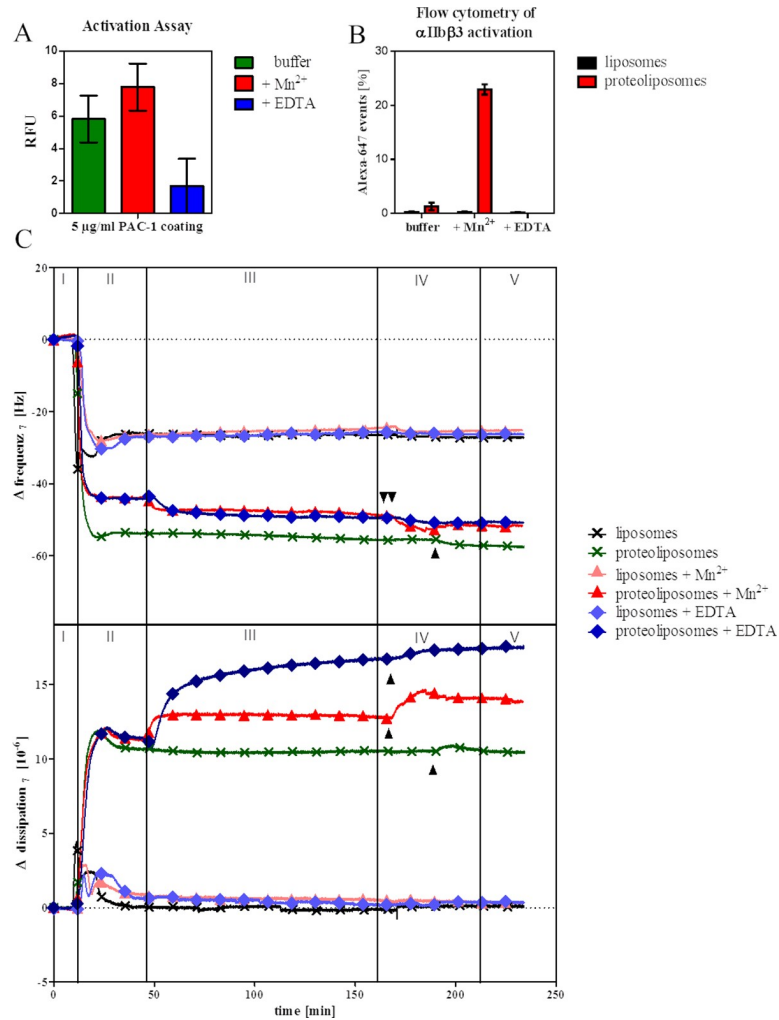

**Fig 3. Activation of αIIbβ3.** A) Activation assay using PAC-1 antibody. Each value is the mean of three replicate measurements ± SEM. The binding of PE CF-liposomes/proteoliposomes to 5 μg/mL PAC-1 coated on a microtiter plate was detected after incubation with buffer (green), 1 mM $Mn^{2+}$ (red) and 5 mM EDTA (blue). Values of bare liposome samples were subtracted from proteoliposome values. The y-axis shows relative fluorescence units (RFU). B) Integrin activation investigated by flow cytometry with PE CF-liposomes (black)/proteoliposomes (red) by adding PAC-1-Alexa 647-coupled antibody. Percentages of the mean ± SEM of Alexa-647 signal of PE positive events incubated with buffer, $Mn^{2+}$ and EDTA are shown. C) Representative QCM-D data showing the changes in frequency $f$ (top) and dissipation $D$ (bottom) of the seventh overtone for the binding of the conformation-specific antibody PAC-1 at 37°C. Buffer was injected over the $SiO_2$ sensors (phase I) and after reaching a baseline liposomes or proteoliposomes were injected and the formation of a bilayer was observed (phase II). After a washing step with either liposome buffer, liposome buffer with 1 mM $Mn^{2+}$, or 5 mM EDTA (phase III), PAC-1 antibody was injected (phase IV) and binding was observed (indicated by arrows). Rinsing with the respective buffer followed (phase V).

show 2% binding of PAC-1, whereas $Mn^{2+}$-treated liposomes show 25% binding (Fig 3B). Additionally, no binding of PAC-1 could be observed in proteoliposomes incubated with EDTA. Furthermore, PAC-1 binding was addressed by the complementary biophysical method quartz crystal microbalance with dissipation monitoring (QCM-D), which detects vesicle rupture and lipid bilayer formation, as well as mass adsorption (frequency $f$) and viscoelastic properties (dissipation $D$) [23]. Fig 3C displays the changes in $f$ ($\Delta f$, top) and $D$ ($\Delta D$, bottom) upon PAC-1 binding to proteoliposomes under different conditions. After a baseline was reached (phase I), samples (liposomes or proteoliposomes) were injected (phase II). Both

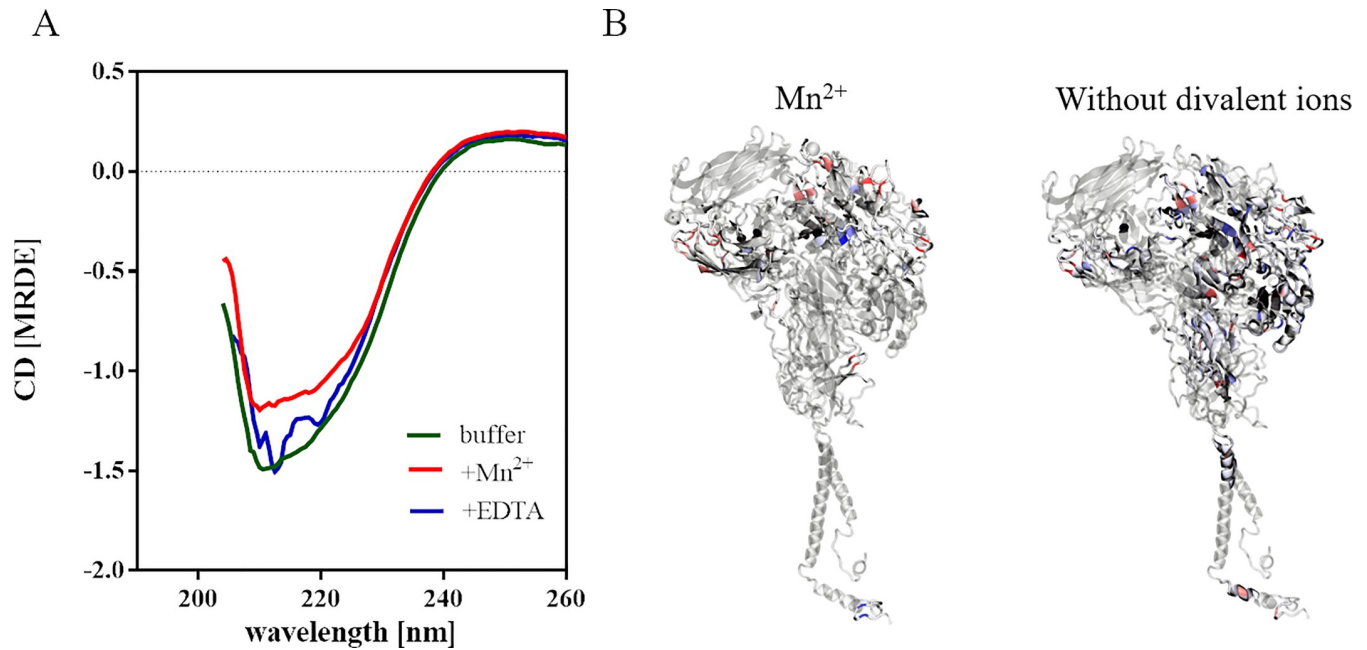

**Fig 4. CD spectra and MDS of αIIbβ3.** A) Far-UV region CD spectra of αIIbβ3 reconstituted into liposomes in buffer (green), with addition of 1 mM $Mn^{2+}$ (red) or 5 mM EDTA (blue). One representative spectrum recorded with proteoliposomes with a protein concentration of approximately 0.4 µM or liposomes in 5 mm path length cuvettes at 37˚C is shown. Liposome spectra were subtracted from the respective proteoliposome spectra. B) MDS demonstrating the regions that changed drastically in the antiparallel β-sheet probability after removing all structural ions (right) or converting the three $Ca^{2+}$ in the MIDAS, ADMIDAS and SyMBS to $Mn^{2+}$ (left). The red and blue colored regions illustrate the formation or loss of β-sheets, respectively, compared to integrin with $Ca^{2+}$ and $Mg^{2+}$ metal ions only. In white or transparent regions, the β-sheet probability did not change.

https://doi.org/10.1371/journal.pone.0214969.g004

bare liposomes and proteoliposomes show a strong binding to the $SiO_2$ substrate as indicated by the decrease in $f$, but only for proteoliposomes, a significant increase in $D$ is observed. The $f$ of pure liposomes (black) reaches a minimum and then increases again due to the release of enclosed aqueous buffer and levels out a stable value of -26 ± 0.2 Hz. The respective $D$ signal shows a response of the opposite direction and reaches a baseline at 0.5 ± 0.1 x $10^{-6}$. However, proteoliposomes (green) display a constant decrease in $f$, that stabilizes at -53 Hz, and an increase in $D$ up to 12 x $10^{-6}$. A schematic illustration of the setup in the flow chambers of the instrument is displayed in S1 Fig. Treatment of the proteoliposome bilayer with buffer containing $Mn^{2+}$ (red) or EDTA (blue) led to a decrease in $f$ of approximately 5 Hz and an increase in $D$ of 2.5 x $10^{-6}$ for $Mn^{2+}$ and 5 x $10^{-6}$ for EDTA due to the buffer injection (phase III), respectively. Subsequently, PAC-1 antibody (phase IV) was injected to determine the activation state of αIIbβ3. The integrin bilayer treated with $Mn^{2+}$ (red) showed a decrease in $f$ of 4 ± 0.3 Hz, whereas for the standard buffer treated bilayer (dark green) PAC-1 binding induces only 1.9 ± 0.7 Hz changes in $f$. EDTA treatment led to even lower changes in $f$. The respective $D$ response is oppositely increasing. By contrast, in control experiments without protein (pink, light blue, black), neither changes of $f$ nor $D$ could be demonstrated in phase IV. Phase V corresponds to rinsing with the respective buffer.

## Changes in αIIbβ3 secondary structure reconstituted into liposomes and their correlation with protein activation

By CD spectroscopy we searched for correlations between the protein secondary structure of αIIbβ3 and its activation state. Fig 4A shows a representative far-UV CD spectrum of proteoliposomes in buffer (green), after treatment with $Mn^{2+}$ (red) and with addition of 5 mM EDTA (blue). After $Mn^{2+}$-treatment, the CD spectrum shows a decrease of amplitude, that can be almost recovered after EDTA treatment.

Complementary to CD measurements, molecular dynamics simulation studies were carried out. All divalent ions were either removed or the three divalent ions in the MIDAS, the adjacent to MIDAS (ADMIDAS) and the synergistic metal ion-binding-side (SyMBS) of the β3-subunit were changed to $Mn^{2+}$. Shifts in the β-sheet probability are illustrated in Fig 4B. Without any divalent ions, several β-sheets in the PSI-domain get destabilized, but new β-sheets are formed in the Calf-1 and 2 domain and existing ones extend in the hybrid domain. β-sheets in the β-propeller region are rearranged. Similar effects are observed after the exchange of the divalent ions in the three metal ion-binding sides comprising again rearrangement in the β-propeller domain, but elongation and formation of new sheets in the Calf-1 and EGF2-domain. The changes in the secondary structure content for both the experimental and the simulated approach are summarized in S1 Table. As detected by CD spectroscopy and supported by MDS results, the α-helical content is reduced by 0.6% and β-sheet content increases by 1% after addition of $Mn^{2+}$. Treatment with EDTA leads to a small decrease in β-sheet content as shown by CD spectroscopy. By contrast, MDS results show an increase of 0.3% in β-sheet content after removal of divalent ions.

### Drug-induced activation of αIIbβ3 and changes in protein secondary structure

QCM-D experiments reveal PAC-1 antibody binding to integrin reconstituted into phospholipid bilayers upon treatment with fondaparinux, UFH or quinine as indicated by changes in $f$ and $D$ (Fig 5A and S2 Fig). The frequency shifts after PAC-1 injection shown in Fig 5A for quinine ($\Delta f = 4.3 \pm 0.3$ Hz) and $Mn^{2+}$ ($4.0 \pm 0.1$ Hz) are comparable and greater than in the absence of drug or $Mn^{2+}$. Fondaparinux and UFH ($\Delta f = 3.0 \pm 0.7$ Hz) do not significantly modify the binding of PAC-1 with respect to the control buffer ($\Delta f = 1.9 \pm 0.7$ Hz), and after complexing the $Mn^{2+}$ by EDTA, only half of the signal is found than with buffer alone.

Results from the activation assay with the PAC-1 antibody coated on a microtiter plate are consistent with these QCM-D data (Fig 5B). The highest binding tendency of PE CF-fluorescently labelled proteoliposomes to PAC-1 is again observed after incubation with quinine or $Mn^{2+}$ compared to the buffer control. UFH-treated proteoliposomes still show more binding of PAC-1 than fondaparinux, which is at the same binding level as the control buffer. Changes in the secondary structure of αIIbβ3 during interaction with the drugs were traced by CD spectroscopy. Fig 5C displays the normalized CD signal as MRDE values at 210 nm wavelength extracted from the far-UV spectra (S3 Fig) of proteoliposomes upon titration with rising concentrations of fondaparinux (purple), UFH (blue) and quinine (black), respectively.

The spectra show an increase in the amplitude with rising concentrations of UFH transforming into decreasing MRDE values at 210 nm (Fig 5C). By contrast, incubation of proteoliposomes with increasing concentration of fondaparinux resulted in insignificant changes of MRDE signal, and no amplitude changes were observed with quinine at concentrations between 0.1–1.5 μg/mL.

Within the experimental error, a linear dependence of the MRDE at 210 nm on the logarithm of the drug concentration, especially UFH, is observed. The concentration dependence of the MRDE value indicates changes in the secondary structure of integrin during drug interaction.

### Discussion

In this study we investigated the correlation between the secondary structure of αIIbβ3 reconstituted into liposomes and its activation state. A membrane-like system (liposomes) was used to exclude any additional interferences with e.g. activators in platelets and successful reconstitution was presented.

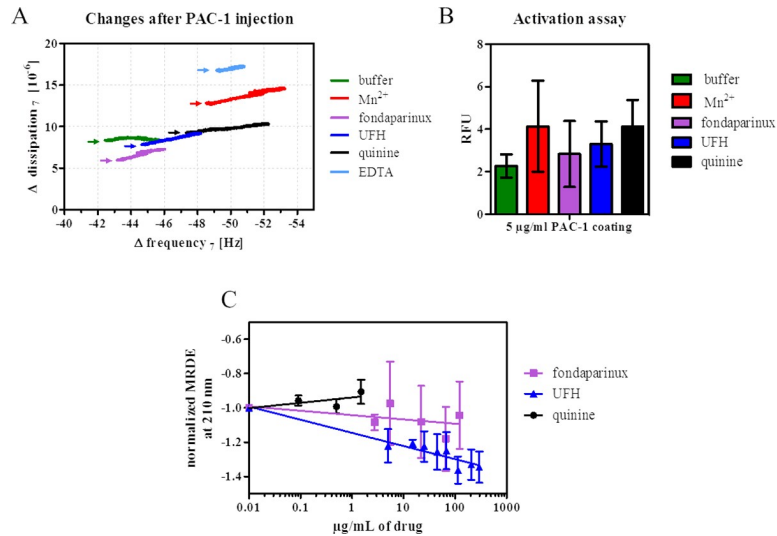

**Fig 5. αIIbβ3 activation by drugs.** A) Changes in frequency (Δf) and dissipation (ΔD) after PAC-1 injection event (indicated by the respective arrows) in representative QCM-D experiments with proteoliposomes after treatment with buffer (green), 1 mM $Mn^{2+}$ (red), 250 μg/mL fondaparinux (purple), 50 μg/mL quinine (black), 250 μg/mL UFH (blue) and 5 mM EDTA (light blue) for at least 15 min at 37˚C. B) Drug activation assay using PAC-1 antibody. Each value is the mean of three replicate measurements ± SEM. The binding of PE CF-liposomes/proteoliposomes to 5 μg/mL PAC-1 coated to a microtiter plate was detected after the incubation with buffer (green), 1 mM $Mn^{2+}$ (red), 250 μg/mL fondaparinux (purple), 250 μg/mL UFH (blue) and 50 μg/mL quinine (black). Liposome sample results were subtracted from proteoliposomes values. The y-axis shows relative fluorescence units (RFU). C) Normalized single wavelength plot for corresponding MRDE values at 210 nm from far-UV region CD spectra of αIIbβ3 incorporated into liposomes/proteoliposomes treated with increasing concentrations of UFH (blue), fondaparinux (purple) and quinine (black), respectively. Normalized averages from three independent measurements ± SEM are shown as dots that were recorded with proteoliposomes with a protein concentration of approximately 0.4 μM or liposomes in 5 mm path length cuvettes at 37˚C. Liposomes spectra were subtracted from the respective proteoliposome spectra. Respective trendlines were applied to guide the reader.

QCM-D measurements enable the analysis of the mechanism of vesicle rupture and formation of supported lipid bilayers [24–26], which makes the analysis of membrane proteins in physiological-like environment and interaction with other molecules feasible [27–29]. Although Frohnmayer *et al.* investigated αIIbβ3 egg PC/egg PG proteoliposomes and its binding to extracellular matrix in QCM-D experiments [30] our setup has not been previously considered for analysis of integrin interactions in a $SiO_2$-supported lipid bilayer. Injection of proteoliposomes show a strong increase in *D* while the decrease in *f* is comparable to the bare liposome measurements. However, the increase in *D* is not as high as it would be observed without any supported lipid bilayer (SLB) formation [31]. The high dissipation was explained by a small amount of liposomes potentially remaining on the assumed substrate-covering SLB and [23, 31] the huge ectodomain of the 235 kDa large integrin (S1 Fig) [32].

QCM-D results, flow cytometry activation detection as well as activation assay demonstrated a higher binding of conformation-specific antibody PAC-1 after treatment with $Mn^{2+}$ than with control buffer and after capturing of divalent ions by EDTA. This indicates an activation of αIIbβ3 by $Mn^{2+}$ and supports the importance of other divalent ions such as $Ca^{2+}$ for the activation process, that have been examined over the past years [11, 30, 33]. Previous MDS studies revealed eight divalent cation binding sites on αIIbβ3. The MIDAS, which is required for ligand binding, is capable for binding $Ca^{2+}$ and $Mg^{2+}$ ions, while the other sites are specific for Ca²⁺. The metal ions induce local structural changes in the MIDAS of integrins, which consequentially leads to a displacement of the α7-helix located in the head region by two helical turns upon the dynamics of αIIbβ3 activation [8, 10]. Although Mn²⁺ ions compete with the divalent ions and increases the ligand affinity, it is controversially discussed whether a complete activation of αIIbβ3 is reached [34]. Our analysis of PAC-1 binding suggests that Mn²⁺ induces indeed an active conformation. The control buffer shows an increased activation of integrin, especially in the activation assay. This indicates a partial activation that could be explained by the long incubation with the PAC-1 antibody or unspecific binding to the microtiter plate leading to activation.

Further, we addressed the question whether the activation/opening of the protein correlates to certain changes in protein secondary structure. Several recent publications identified the structure for parts of other integrin subunits by CD spectroscopy e.g. the domain I of CD11b/CD18 leukocyte integrin [35] or the biological active part of α3β1 integrin [36] and even the cytoplasmic tail and the transmembrane domain of αIIbβ3 [37]. Additionally, MDS studies as well as crystal structures have shown the dynamics of integrin motion [38, 39]. However, these studies concentrated only on specific domains of proteins and/or have not taken into account the dynamic rearrangements during activation of overall integrin within a lipid environment.

The evaluation of CD spectra from proteoliposomes is complex due to solvent shifts caused by hydrophobic lipid environment, differential light scattering or absorption flattening effects [40]. Moreover, data quality depends strongly on the lipid-to-protein ratio, which was in our study as high as about 1000:1 [40, 41]. Therefore, lower wavelengths could not be included into the evaluation of the shown CD spectra and the data could be deconvoluted only to a limited extent. The CD results support the switchblade model for integrin activation shown by various groups [6, 12, 42], where a dramatic rearrangement of the global structure by a switchblade-like opening of the interface between the headpiece and stalk occurs, extending the ligand binding domain away from the plasma membrane. Nevertheless, major changes in secondary structure upon activation have not been shown in previous studies as well as in our experimental approach, which is explained by probable motion of mainly disordered regions e.g. the flexible knee region between the EGF domains of the β-subunit [6].

The MDS results support minor changes in the β-sheet content of the knee region of αIIb- and β3-subunit (the Calf-1 and EGF2 region, respectively), that are only visible by localizing the overall β-sheet probability changes. This could lead to a decrease in the interaction of Thigh↔Calf1 and EGF1↔EGF2, and eventually reduce the force needed to unfold the protein (e.g. in a switchblade motion after Mn²⁺-treatment). The changes in this region could get conveyed from the MIDAS site by the rearrangements of β-sheets in the β-propeller domain. With the removal of all divalent ions the secondary structure changes were stronger and almost random. They included changes in all extracellular αIIb domains. In the simulation was assumed, that EDTA coordinates and removes all divalent ions. In the experiment, certain ions might not be accessible by EDTA and remain coordinated to the integrin. This may explain the contrast in the results between MDS and experiments. If assumed that only the ions in MIDAS, AMIDAS and SyMDS are removed, the changes in protein secondary structure might be neglectable as observed in the experiments.

Our methods were applied to demonstrate that clinically relevant drugs quinine, UFH and fondaparinux have an impact on activation of αIIbβ3. Especially, the quinine-treated integrin leads to PAC-1 binding indicating the activation and opening of αIIbβ3. However, no changes in secondary structure were observed in CD spectroscopy upon titration with low concentrations of quinine.

Due to its chirality, quinine has an influence on the CD spectra, and the plasma concentration of quinine in uncomplicated falciparum malaria patients of approx. 6 μg/mL [43] was not attained in these measurements. The potential binding site of quinine to αIIbβ3 is involved in the transition from the bent to the active form of the integrin by a swing-out motion because of the β3 A-α7-helix movement towards the hybrid domain [8]. Therefore, potential binding of quinine to this domain could lead to changes in the protein structure or even to integrin activation. Two explanations are possible for the interaction and activation of αIIbβ3 with quinine. Firstly, quinine binds to αIIbβ3 and leads to an extension of the head domain (activation), which enables PAC-1 binding. Secondly, quinine mediates the contact between the PAC-1 antibody and αIIbβ3 independently from the activation state and enhances the affinity of this interaction. This was described recently for mouse quinine-dependent antibodies [44–46]. We assume that the interaction of quinine with αIIbβ3 activates the integrin and induces the exposure of cryptic epitopes, e.g. ligand-induced binding sites, which could lead to drug-induced thrombocytopenia [17].

Treatments with UFH and with $Mn^{2+}$ result in opposite influence on the CD-amplitude, but both lead to increased binding of PAC-1 in QCM-D. Possibly, $Mn^{2+}$ induces other changes in secondary structure than UFH, and activation of αIIbβ3 is achieved in both cases *via* different mechanisms. Additionally, activation assays indicated rising amounts of activated αIIbβ3 upon UFH treatment. Studies with other integrins, e.g. αX β2 [47] and αV β3 [48], showed heparin binding near the ligand-binding site in the integrin head domain. These regions are homologous to e.g. the A-ligand binding domain of the β3-subunit. Binding of UFH to the region containing the RGD binding pocket could lead to opening of the integrin mediating binding of PAC-1 [21, 47, 48]. By contrast, incubation of αIIbβ3 with fondaparinux induces neither significant activation of the integrin, nor changes in its secondary structure. Apparently, sulfated polysaccharides with different lengths interact differently with αIIbβ3, similarily to other proteins [49]. Besides the mechanism of platelet-factor 4/heparin complexes in heparin-induced thrombocytopenia, it is likely that heparin potentiates platelet activation *via* binding to αIIbβ3 leading to thrombocytopenia [19].

Our results reveal that similarly to $Mn^{2+}$, quinine- and UFH induce αIIbβ3 activation in a membrane-like environment, but do not lead to significant changes in the protein secondary structure. This is consistent with switchblade motion that enables the movement of the extracellular parts of αIIbβ3. The exposure of these hidden regions has important clinical implications, because autoantibody binding leads to loss of self-tolerance and immunogenicity in some patients [16, 17]. Our experimental setup could be applied in prospective experiments to reveal autoantibody binding to αIIbβ3 under various conditions. In addition, the combination of the biophysical and biological tools can be applied in future to study other transmembrane proteins.

## Materials and methods

### Proteins and chemicals

Human integrin αIIbβ3 was purchased from Enzyme Research Laboratories (South Bend, USA). Unless otherwise stated, all mentioned antibodies were bought from Biolegend (San Diego, USA). Dimyristoylphosphatidylglycerol (DMPG; 14:0 PG), dimyristolphosphatidylcholine (DMPC; 14:0 PC) and dioleoyl-glycero-phosphoethanolamine-N-carboxyfluorescein (PE CF) were obtained from Avanti Polar Lipids Inc. (Alabaster, USA). SM-2 biobeads were supplied by Bio-Rad (Munich, Germany). Unfractionated heparin (UFH), fondaparinux and quinine sulfate were purchased from Sigma-Aldrich (Steinheim, Germany). Protein concentration was determined by bicinchonic acid assay (BCA) kit with included protein microstandard (Sigma-Aldrich, Steinheim, Germany). Tris-Base, ethylenediaminetetraacetic acid (EDTA), bovine serum albumin and NaCl were bought from Sigma-Aldrich (Tauf-kirchen, Germany). $CaCl_2$, $MnCl_2$, Triton X-100 and methanol were purchased from Carl Roth GmbH (Karlsruhe, Germany). Sucrose and sodium dodecyl sulfate (SDS) were obtained from Merck KgaA (Darmstadt, Germany).

## Reconstitution of integrin αIIbβ3 into liposomes

Liposome buffer consisting of 20 mM Tris, 50 mM NaCl, 1 mM $CaCl_2$ was prepared and the pH was adjusted to 7.4 with HCl. Liposomes were prepared following an adapted protocol of Erb and Engel [50]. Briefly, a 900 nM DMPG:DMPC (1:20) mixture was dried first under nitrogen stream and then in vacuum overnight. For the activity assays, fluorescein labeled lipids (PE CF) were added to the lipid mixture with a ratio of DMPC:DMPG:PE CF 500:25:1. Subsequently, lipids were dissolved in liposome buffer containing 0.1% Triton X-100 and 0.2 mg/mL integrin αIIbβ3 (1000:1 lipid:protein molar ratio). The solution was incubated for 2 h at 37°C and Triton X-100 was removed afterwards by adding twice 50 mg SM-2 biobeads for 210 min at 37°C. The biobeads were prewashed with methanol and ultrapure water (Sartorius, Göttingen, Germany). Non-reconstituted αIIbβ3 was separated from the proteoliposomes by ultracentrifugation at 4°C and 268,000 $g$ for 24 h with a four-step sucrose gradient (2 M, 1.2 M, 0.8 M and 0.4 M in liposome buffer). The proteoliposome-containing fraction was collected and dialyzed against liposome buffer for 72 h using 8 kDa cut-off dialysis cassettes (GE Health-care, Freiburg, Germany). These vesicles were stored at 4°C and used for experiments within four days.

## Validation of protein reconstitution into liposomes

**Determination of protein concentration.** Determination of protein concentration was done by a modified BCA adapted from Parmar et al. [51]. Protein microstandards or 25 μL proteoliposome samples were pipetted into a microtiter plate (Sarstedt AG, Nürnbrecht, Germany). After addition of 25 μL 0.5% SDS, 50 μL of working BCA reagent was added to each well. The plate was incubated for 2 h at 37°C and the absorbance was measured at 562 nm on a TECAN infinite M200Pro plate reader (Tecan group, Männedorf, Swizerland). The protein concentration in the samples was determined from a standard curve with the mentioned protein standard.

**SDS-PAGE.** SDS-PAGE was carried out to detect the integrin subunits under reductive conditions upon protein reconstitution into liposomes. Liposome and integrin liposome samples were loaded to 4–12% gradient Bis-Tris SDS gels and visualized with the Pierce Silver Stain Kit (both Thermo Fischer, Darmstadt, Germany).

**Transmission electron microscopy (TEM) of negatively stained proteoliposomes.** For the negative staining procedure, the proteoliposomes were allowed to adsorb onto a glow-discharged carbon-coated holey Pioloform film on a 400-mesh grid for 5 min. The grid was then transferred onto two droplets of deionized water and then onto a droplet of 2% aqueous phosphotungstic acid at pH 7.0 (VWR, Radnor, USA) for 30 s and finally on a second droplet for 4 min. After blotting with filter paper and air-drying, the samples were examined with a transmission electron microscope LEO 906 (Carl Zeiss Microscopy GmbH, Oberkochen, Germany) at an acceleration voltage of 80 kV. Pictures were taken with flat films with a magnification of 60,000-times. Afterwards, the micrographs were analyzed using Adobe Photoshop CS6.

**Dynamic light scattering (DLS) measurements.** Dynamic light scattering (DLS) measurements were acquired on a Zetasizer Nano ZS (Malvern Instruments, Herrenberg, Germany). Liposome and proteoliposome samples were diluted in liposome buffer (1:10) and vacuum degassed for 20 min at 37˚C. Subsequently, 500 µl were transferred to a 10 mm path length cuvette (Brand, Wertheim, Germany) and equilibrated 4 min at 37˚C. Measurements were done at a detector angle of 173˚ with a refractive index of 1.45 and absorption of 0.001 with standard solvent parameters as referred to water. Each measurement consisted of 15 runs and was repeated five times. Hydrodynamic diameter data was analyzed with the Zetasizer software 7.11.

**Flow cytometry.** Flow cytometry experiments were carried out for the validation of successful integrin reconstitution into liposomes. Integrin was detected by anti-CD61-Alexa 647 and anti-CD41 (Bio-Techne Holding, Minneapolis, USA) antibody binding. For analysis, 10 µl of purified PE CF-fluorescently labelled liposomes were mixed with 1 µL of 200 µg/mL anti-CD61-Alexa 647 and 1 µL of 500 µg/mL anti-CD41 and incubated for 30 min at room temperature (RT). The anti-CD41 antibody samples were then incubated with 1 µL of the secondary goat anti-mouse IgG-Alexa Fluor 647 (Thermo Fischer) antibody for 10 min at RT. Afterwards, samples were diluted with 290 µl liposome buffer and analyzed in a BD LSR II Flow Cytometer (Becton, Dickinson and Company; Franklin lakes, USA). Integrin activation was detected by adding 1 µL of 100 mg/mL conformation-specific Alexa 647-labelled PAC-1 IgM antibody to PE CF-liposomes in buffer, with 1 mM $MnCl_2$ (control for activation) or EDTA (negative control). The FlowJo 7.6.5 software was used for evaluation.

## Activation assay for proteoliposomes

Activation assay for proteoliposomes was carried out following a protocol adapted from Ye et al. 2009 [7]. PAC-1 was coated on a 96 well plate (Capitol Scientific, Austin, USA) at 4˚C overnight. After blocking the plate with liposome buffer containing 30 mg/mL bovine serum albumin, PE CF-fluorescently labelled liposomes or proteoliposomes were added (conditions are indicated in the graphs) and incubated for 4 h at 37˚C. Unbound liposomes were washed away with liposome buffer and after addition of 100 µL 1% Triton- X-100 in liposome buffer, fluorescence was read in a microplate reader (Paradigm, Beckman Coulter, Pasadena, USA) at ex485/em535 nm. The respective liposome fluorescence signal was subtracted from the proteoliposome signal.

## Quartz crystal microbalance with dissipation monitoring (QCM-D)

QCM-D measurements were carried out with a Q-sense Analyzer from Biolin Scientific Holding AB (Västra Frölunda, Sweden) under continuous flow of 25 µL/min driven by a peristaltic pump (Ismatec IPC-N4, Idex Health & Science GmbH, Wertheim-Mondfeld, Germany) at 37˚C. $SiO_2$-coated quartz crystal sensors (Biolin Scientific) were cleaned using a 2% SDS solution for 30 min at RT followed by rinsing with ultrapure water. Afterwards, crystals were dried under a stream of nitrogen and exposed for 20 min to UV-ozone (Pro Cleaner Plus, Bioforce Nanoscience, Ames, USA). Resonance frequency and dissipation were measured at several harmonics (15, 25, 35, 45, 55, 65 MHz) simultaneously. Changes in dissipation ($\Delta D$) and frequency ($\Delta f$) of the seventh overtone (35 MHz) are presented in the graphs. After equilibrating the system with liposome buffer (15 min), liposomes or proteoliposomes were injected into the system. After surface adsorption, the system was washed for at least 20 min with liposome buffer. For the activation state experiments, liposome buffer containing 1 mM $MnCl_2$ or 5 mM EDTA was loaded into the system and incubated under continous flow for approximately 1 h. PAC-1 was finally added at a concentration of 5 µg/mL for interaction analysis followed by rinsing with the respective buffer. For drug interaction analyses, 250 µg/mL UFH, 250 µg/mL fondaparinux or 50 µg/mL quinine sulfate diluted in liposome buffer were introduced into the system for 10 min after washing and PAC-1 injection was performed subsequently. Data analysis was achieved using Q-Tools V.3.0 and QSoft401 V2.5 (both Biolin Scientific AB).

## Circular dichroism (CD) spectroscopy

CD spectra were measured with a Chirascan CD spectrometer (Applied Photophysics, Leatherhead, UK) equipped with a temperature control unit (Quantum Northwest, Liberty Lake, USA) at 37°C. Measurements at wavelengths in the range of 195–360 nm were performed with a 5 mm path length cuvette (110-QS; Hellma Analytics, Müllheim, Germany) and a protein concentration of 0.4 μM incorporated into liposomes. Spectra were recorded with a bandwidth of 1.0 nm, a scanning speed of 15 nm/min and three repetitions. For data analysis, spectra of liposomes were subtracted from proteoliposomes. For the analysis of $Mn^{2+}$ and EDTA induced changes in the secondary structure of αIIbβ3, the sample was first measured in liposome buffer, afterwards $MnCl_2$ was added to a final concentration of 1 mM and the sample was incubated 45 min before measurement. Finally, 50 mM EDTA was added to the same cuvette to a final concentration of 5 mM and incubated for 45 min at RT. For the drug titrations, respective volume of UFH (0–295 μg/mL), fondaparinux (0–122 μg/mL) or quinine sulfate (0–1.5 g/mL) were added to the cuvette and equilibrated for 5 min at 37°C before each measurement. Normalization of the data was calculated by the wavelength dependent mean residue delta epsilon (MRDE) that includes concentration $c$, number of amino acids $AA$, and path length of the cuvette $d$:

$$MRDE = \frac{CD[mdeg]}{c[M] \cdot AA \cdot d[cm] \cdot 32.982}$$

The estimation of the secondary structural content was done by deconvolution of CD spectra by the CDNN software using a database of 33 reference proteins [52].

## Molecular dynamics simulations (MDS)

**Integrin model.** For the complete bent integrin model, the NMR structure of the transmembrane domain (PDB-code: 2k9j) [53] and the X-ray structure of the ectodomain (PDB-code: 3fcs) [6] were combined. The missing residues 764 to 774 and 840 to 873 in the αIIb chain, and 75 to 78 and 477 to 482 in the β3 chain were added as random coil. After manual linking of the fragments, the structure was energy minimized and equilibrated in vacuum to relax overstretched bonds. The glycosylations and modified residues as described in the UniProt entries (P08514 and P05106) [54] were considered. A detailed list is provided in the Supporting Information (S2 Table). Seven $Ca^{2+}$ ions and one $Mg^{2+}$ ion, as present in the X-ray structure, were inherited. The integrin was then embedded in a membrane containing 1076 DMPC and 54 DMPG lipids (20:1), solvated and equilibrated for 350 ns in a isothermal-isobaric (NPT) ensemble to ensure that the created gap in the membrane at the integrin membrane interface is closed. Ions were added to neutralize the box and the ion content was not adjusted to physiological conditions. For a more detailed description of the simulation cell, please refer to the Supporting information S3 Table. In the time scale of several hundreds ns the activation and opening of integrin cannot be achieved without external forces during a MDS. Therefore, an open conformation of integrin was obtained with steered MD by constant velocity pulling. The transmembrane domain and the integrin head (β-propeller and β3 A-domain) were pulled apart from each other with a speed of 2 m/s up to complete unfolding. The real distance was restrained to the constantly increasing virtual group distance by a harmonic potential with a force constant of 50 kJ·mol$^{-1}$·nm$^{-2}$.

**Simulation parameters.** The equation of motion was integrated with a Verlet integrator applied every 7 fs with the simulation software package GROMACS 5.1 [55]. This was achieved by increasing the mass of all hydrogen to 4 m.u., and reducing the corresponding heavy atom mass by the same amount (hydrogen mass repartitioning—HMR). Explicit water was described by the TIP3P model [56], the protein with the AMBER99SB-ildn force field [57], modified amino acids with the PTM force field [58], carbohydrates with the Glycam force field [59] and lipids with the Slipids force field [60]. Periodic boundary conditions were used. The intermolecular interactions were applied with the particle mesh Ewald (PME) method [61] and a grid spacing of 0.12 nm for Coulomb interactions, and van der Waals interactions with Lennard-Jones functions. Cutoffs of 1.0 nm with switching functions at 0.9 nm were used with a neighborlist distance [62] of 1.2 nm. All bonds were constrained with the LINCS algorithm [63] to the optimal distance. The temperature bath was regulated to 300 K with a modified v-rescale thermostat [64] for protein and non-protein separately every 100 fs. The pressure was controlled to 1 bar every 12 ps by a Parrinello-Rahman barostat [65]. The barostat is semiisotropic, and relaxation of the membrane in xy-direction and control of the water density by adjusting the height in z-direction are independent of each other.

**MDS of the effect of ions on the integrin structure.** Three systems with membrane embedded integrin, water and different structural ion compositions were prepared: i) Seven $Ca^{2+}$ and one $Mg^{2+}$ ion as present in the X-ray structure [6]; ii) the three structural ions in the proximity of the active center were replaced with $Mn^{2+}$; and iii) all structure ions were removed. Each system was first equilibrated for 350 ns in the NPT, followed by 100 ns production trajectory, and conformations were collected every 700 ps. The structural distribution and MDS secondary structure (S1 Table) were analyzed and predicted with the AmberTools program CppTraj [66] over the whole trajectory.

## Supporting information

**S1 Fig. Schematic illustration of the formation of a proteoliposome-derived bilayer with reconstituted αIIbβ3 on a SiO₂ quartz sensor.** We assume in our QCM-D experiments the fusion of proteoliposomes to the substrate and the formation of a bilayer. Potentially, some proteoliposomes remain on the lipid bilayer surface. Conformation specific antibody PAC-1 (cyan) could bind to activated αIIbβ3 in a bilayer as well as to activated αIIbβ3 in liposomes. (TIF)

**S2 Fig. Representative QCM-D data showing the changes in frequency $\Delta f$ (top) and dissipation $\Delta D$ (bottom) of the seventh overtone for the binding of PAC-1 antibody at 37˚C.** Buffer was injected over the SiO₂ sensors and after reaching a baseline, liposomes or proteoliposomes were injected and formation of a bilayer was observed. After a washing step with buffer, the bilayer was treated with the respective drugs (250 μg/mL fondaparinux, 250 μg/mL UFH and 50 μg/mL quinine sulfate), which is indicated by the first arrow and PAC-1 antibody was injected (indicated by the second arrow) followed by rinsing with the respective buffer. (TIF)

**S3 Fig. The far-UV region CD spectra of αIIbβ3 reconstituted into liposomes titrated with drugs.** The far-UV region CD spectra of αIIbβ3 reconstituted into liposomes in buffer (dark green), and with increasing concentrations of UFH (top), fondaparinux (middle) and quinine (bottom), respectively. Representative spectra recorded for proteoliposomes with a protein concentration of approximately 0.4 μM in 5 mm path length cuvettes at 37˚C are shown. Liposome spectra were subtracted from the respective proteoliposome spectra. (TIF)

**S1 Table. Change in the secondary structure distribution of αIIbβ3 determined by CD spectroscopy and MDS.** Changes in the secondary structure distribution between integrin αIIbβ3 in buffer environment, after addition of 1 mM $Mn^{2+}$ experimentally or changing the three ions in the MIDAS and ADMIDAS region to $Mn^{2+}$ *via* MDS, and after addition of 5 mM

EDTA experimentally or removing all structural ions during MDS in the environment. The estimation of the experimental secondary structure content was carried out with the deconvolution of CD spectra using CDNN software. The MDS secondary structure was predicted with CPPTRAJ.
(TIF)

**S2 Table. Amino acid modifications of the MDS model of αIIbβ3.** Amino acid modifications of the MDS model of integrin αIIbβ3. NDG and NAG are N-Acetylglucosamine in α and β form, respectively, MAN is α-Mannose and NGA is β-N-Acetylgalactosamine.
(TIF)

**S3 Table. System setups for the particular MDS.** System setups for the particular MDS. The first five columns indicate the cell volume Cs, the number of water molecules $N_W$, membrane molecules $N_{DMPC}$ and $N_{DMPG}$, and ions $N_I$. The last two columns contain the equilibration $t_E$ and data collection $t_D$ times.
(TIF)

## Acknowledgments

We acknowledge Eric Tönnies for technical support with CD measurements, Felix Nagel for the help with the generation of the PyMOL protein structures, Doreen Biedenweg for assistance with the FACS measurements, Dr. Rabea Schlüter for TEM imaging, and Peter Nestler for fruitful discussions. We further thank the BRAIN supercomputing cluster of the University of Greifswald for providing the necessary computational resources.

## Author Contributions

**Conceptualization:** Una Janke, Mihaela Delcea.

**Data curation:** Una Janke, Martin Kulke.

**Formal analysis:** Una Janke, Martin Kulke.

**Funding acquisition:** Mihaela Delcea.

**Investigation:** Una Janke, Mihaela Delcea.

**Methodology:** Una Janke, Martin Kulke.

**Project administration:** Mihaela Delcea.

**Resources:** Mihaela Delcea.

**Supervision:** Walter Langel, Mihaela Delcea.

**Validation:** Una Janke, Martin Kulke, Ina Buchholz, Norman Geist, Walter Langel.

**Visualization:** Una Janke.

**Writing – original draft:** Una Janke, Mihaela Delcea.

**Writing – review & editing:** Una Janke, Martin Kulke, Ina Buchholz, Norman Geist, Walter Langel, Mihaela Delcea.

## References

1. Bledzka K, Pesho MM, Ma Y-Q, Plow EF. Chapter 12—Integrin αIIbβ3. In: Michelson AD, editor. Platelets (Third Edition). Third Edition ed: Academic Press; 2013. p. 233–48.

2. Ma YQ, Qin J, Plow EF. Platelet integrin alpha(IIb)beta(3): activation mechanisms. Journal of thrombosis and haemostasis: JTH. 2007; 5(7):1345–52. https://doi.org/10.1111/j.1538-7836.2007.02537.x PMID: 17635696.

3. Shattil SJ, Newman PJ. Integrins: dynamic scaffolds for adhesion and signaling in platelets. Blood. 2004; 104(6):1606–15. https://doi.org/10.1182/blood-2004-04-1257 WOS:000223818600012. PMID: 15205259

4. Litjens PE, Akkerman JW, van Willigen G. Platelet integrin alphaIIbbeta3: target and generator of signalling. Platelets. 2000; 11(6):310–9. PMID: 11083455.

5. Bennett JS. Structure and function of the platelet integrin alphaIIbbeta3. The Journal of clinical investigation. 2005; 115(12):3363–9. https://doi.org/10.1172/JCI26989 PMID: 16322781; PubMed Central PMCID: PMC1297263.

6. Zhu J, Luo BH, Xiao T, Zhang C, Nishida N, Springer TA. Structure of a complete integrin ectodomain in a physiologic resting state and activation and deactivation by applied forces. Molecular cell. 2008; 32 (6):849–61. https://doi.org/10.1016/j.molcel.2008.11.018 PMID: 19111664; PubMed Central PMCID: PMC2758073.

7. Ye F, Liu J, Winkler H, Taylor KA. Integrin alpha IIb beta 3 in a membrane environment remains the same height after Mn2+ activation when observed by cryoelectron tomography. Journal of molecular biology. 2008; 378(5):976–86. https://doi.org/10.1016/j.jmb.2008.03.014 PMID: 18405917; PubMed Central PMCID: PMC2614134.

8. Campbell ID, Humphries MJ. Integrin Structure, Activation, and Interactions. Csh Perspect Biol. 2011; 3 (3).ARTN a004994 https://doi.org/10.1101/cshperspect.a004994 WOS:000287846200012.

9. Xu XP, Kim E, Swift M, Smith JW, Volkmann N, Hanein D. Three-Dimensional Structures of Full-Length, Membrane-Embedded Human alpha(IIb)beta(3) Integrin Complexes. Biophysical journal. 2016; 110 (4):798–809. https://doi.org/10.1016/j.bpj.2016.01.016 PMID: 26910421; PubMed Central PMCID: PMC4776043.

10. Zhang K, Chen JF. The regulation of integrin function by divalent cations. Cell Adhes Migr. 2012; 6 (1):20–9. https://doi.org/10.4161/cam.18702 WOS:000304611500005. PMID: 22647937

11. Ye F, Kim C, Ginsberg MH. Reconstruction of integrin activation. Blood. 2012; 119(1):26–33. https://doi.org/10.1182/blood-2011-04-292128 PMID: 21921044; PubMed Central PMCID: PMC3251231.

12. Takagi J, Petre BM, Walz T, Springer TA. Global conformational rearrangements in integrin extracellular domains in outside-in and inside-out signaling. Cell. 2002; 110(5):599–11. PMID: 12230977.

13. Taub R, Gould RJ, Garsky VM, Ciccarone TM, Hoxie J, Friedman PA, et al. A monoclonal antibody against the platelet fibrinogen receptor contains a sequence that mimics a receptor recognition domain in fibrinogen. The Journal of biological chemistry. 1989; 264(1):259–65. PMID: 2909518.

14. Xiao T, Takagi J, Coller BS, Wang JH, Springer TA. Structural basis for allostery in integrins and binding to fibrinogen-mimetic therapeutics. Nature. 2004; 432(7013):59–67. https://doi.org/10.1038/nature02976 PMID: 15378069; PubMed Central PMCID: PMC4372090.

15. Shattil SJ, Hoxie JA, Cunningham M, Brass LF. Changes in the Platelet Membrane Glycoprotein-IIb-IIia Complex during Platelet Activation. J Biol Chem. 1985; 260(20):1107–14. WOS:A1985AQQ3400041.

16. Bhoria P, Sharma S, Varma N, Malhotra P, Varma S, Luthra-Guptasarma M. Effect of steroids on the activation status of platelets in patients with Immune thrombocytopenia (ITP). Platelets. 2015; 26 (2):119–26. https://doi.org/10.3109/09537104.2014.888546 WOS:000351740700003. PMID: 24617442

17. Visentin GP, Liu CY. Drug-induced thrombocytopenia. Hematology/oncology clinics of North America. 2007; 21(4):685–96, vi. https://doi.org/10.1016/j.hoc.2007.06.005 PMID: 17666285; PubMed Central PMCID: PMC1993236.

18. Yagi M, Murray J, Strand K, Blystone S, Interlandi G, Suda Y, et al. Heparin modulates the conformation and signaling of platelet integrin alpha IIb beta 3. Thromb Res. 2012; 129(6):743–9. https://doi.org/10.1016/j.thromres.2011.11.054 WOS:000304263100013. PMID: 22197178

19. Gao C, Boylan B, Fang J, Wilcox DA, Newman DK, Newman PJ. Heparin promotes platelet responsiveness by potentiating alphaIIbbeta3-mediated outside-in signaling. Blood. 2011; 117(18):4946–52. https://doi.org/10.1182/blood-2010-09-307751 PMID: 21368290; PubMed Central PMCID: PMC3100701.

20. Xiao Z, Theroux P. Platelet activation with unfractionated heparin at therapeutic concentrations and comparisons with a low-molecular-weight heparin and with a direct thrombin inhibitor. Circulation. 1998; 97(3):251–6. PMID: 9462526.

21. Hashemzadeh M, Furukawa M, Goldsberry S, Movahed MR. Chemical structures and mode of action of intravenous glycoprotein IIb/IIIa receptor blockers: A review. Experimental and clinical cardiology. 2008; 13(4):192–7. PMID: 19343166; PubMed Central PMCID: PMC2663484.

22. Hantgan RR, Rocco M, Nagaswami C, Weisel JW. Binding of a fibrinogen mimetic stabilizes integrin alpha IIb beta 3's open conformation. Protein Sci. 2001; 10(8):1614–26. https://doi.org/10.1110/ps.3001 PMID: 11468358

23. Richter R, Mukhopadhyay A, Brisson A. Pathways of lipid vesicle deposition on solid surfaces: a combined QCM-D and AFM study. Biophysical journal. 2003; 85(5):3035–47. https://doi.org/10.1016/S0006-3495(03)74722-5 PMID: 14581204; PubMed Central PMCID: PMC1303580.

24. Jing Y, Trefna H, Persson M, Kasemo B, Svedhem S. Formation of supported lipid bilayers on silica: relation to lipid phase transition temperature and liposome size. Soft matter. 2014; 10(1):187–95. https://doi.org/10.1039/c3sm50947h PMID: 24651504.

25. Hardy GJ, Nayak R, Zauscher S. Model cell membranes: Techniques to form complex biomimetic supported lipid bilayers via vesicle fusion. Current opinion in colloid & interface science. 2013; 18(5):448–58. https://doi.org/10.1016/j.cocis.2013.06.004 PMID: 24031164; PubMed Central PMCID: PMC3767439.

26. Lind TK, Cardenas M. Understanding the formation of supported lipid bilayers via vesicle fusion-A case that exemplifies the need for the complementary method approach (Review). Biointerphases. 2016; 11 (2).Artn 020801 https://doi.org/10.1116/1.4944830 WOS:000379582700029

27. Inci F, Celik U, Turken B, Ozer HO, Kok FN. Construction of P-glycoprotein incorporated tethered lipid bilayer membranes. Biochemistry and biophysics reports. 2015; 2:115–22. https://doi.org/10.1016/j.bbrep.2015.05.012 PMID: 29124152; PubMed Central PMCID: PMC5668657.

28. Li X, Wang R, Wicaksana F, Zhao Y, Tang C, Torres J, et al. Fusion behaviour of aquaporin Z incorporated proteoliposomes investigated by quartz crystal microbalance with dissipation (QCM-D). Colloids and surfaces B, Biointerfaces. 2013; 111:446–52. https://doi.org/10.1016/j.colsurfb.2013.06.008 PMID: 23850749.

29. Graneli A, Rydstrom J, Kasemo B, Hook F. Formation of supported lipid bilayer membranes on SiO2 from proteoliposomes containing transmembrane proteins. Langmuir. 2003; 19(3):842–50. https://doi.org/10.1021/la026231w WOS:000180737100056.

30. Frohnmayer JP, Bruggemann D, Eberhard C, Neubauer S, Mollenhauer C, Boehm H, et al. Minimal synthetic cells to study integrin-mediated adhesion. Angew Chem Int Ed Engl. 2015; 54(42):12472–8. https://doi.org/10.1002/anie.201503184 PMID: 26257266; PubMed Central PMCID: PMC4675076.

31. Richter RP, Berat R, Brisson AR. Formation of solid-supported lipid bilayers: an integrated view. Langmuir. 2006; 22(8):3497–505. https://doi.org/10.1021/la052687c PMID: 16584220.

32. Adair BD, Yeager M. Three-dimensional model of the human platelet integrin alpha IIbbeta 3 based on electron cryomicroscopy and x-ray crystallography. Proceedings of the National Academy of Sciences of the United States of America. 2002; 99(22):14059–64. https://doi.org/10.1073/pnas.212498199 PMID: 12388784; PubMed Central PMCID: PMC137836.

33. Litvinov RI, Bennett JS, Weisel JW, Shuman H. Multi-step fibrinogen binding to the integrin (alpha)IIb (beta)3 detected using force spectroscopy. Biophysical journal. 2005; 89(4):2824–34. https://doi.org/10.1529/biophysj.105.061887 PMID: 16040750; PubMed Central PMCID: PMC1366781.

34. Kamata T, Handa M, Sato Y, Ikeda Y, Aiso S. Membrane-proximal alpha/beta stalk interactions differentially regulate integrin activation. J Biol Chem. 2005; 280(26):24775–83. https://doi.org/10.1074/jbc.M409548200 WOS:000230114000060. PMID: 15863495

35. Fairbanks MB, Pollock JR, Prairie MD, Scahill TA, Baczynskyj L, Heinrikson RL, et al. Purification and structural characterization of the CD11b/CD18 integrin alpha subunit I domain reveals a folded conformation in solution. FEBS letters. 1995; 369(2–3):197–201. PMID: 7649257.

36. Furrer J, Luy B, Basrur V, Roberts DD, Barchi JJ Jr. Conformational analysis of an alpha3beta1 integrin-binding peptide from thrombospondin-1: implications for antiangiogenic drug design. Journal of medicinal chemistry. 2006; 49(21):6324–33. https://doi.org/10.1021/jm060833l PMID: 17034138.

37. Li R, Babu CR, Valentine K, Lear JD, Wand AJ, Bennett JS, et al. Characterization of the monomeric form of the transmembrane and cytoplasmic domains of the integrin beta 3 subunit by NMR spectroscopy. Biochemistry. 2002; 41(52):15618–24. PMID: 12501190.

38. Puklin-Faucher E, Gao M, Schulten K, Vogel V. How the headpiece hinge angle is opened: new insights into the dynamics of integrin activation. Journal of Cell Biology. 2006; 175(2):349–60. https://doi.org/10.1083/jcb.200602071 WOS:000241575500016. PMID: 17060501

39. Gahmberg CG, Fagerholm SC, Nurmi SM, Chavakis T, Marchesan S, Gronholm M. Regulation of integrin activity and signalling. Bba-Gen Subjects. 2009; 1790(6):431–44. https://doi.org/10.1016/j.bbagen.2009.03.007 WOS:000267191700006. PMID: 19289150

40. Miles AJ, Wallace BA. Circular dichroism spectroscopy of membrane proteins. Chemical Society reviews. 2016; 45(18):4859–72. https://doi.org/10.1039/c5cs00084j PMID: 27347568.

41. Ciancaglini P, Simao AMS, Bolean M, Millan JL, Rigos CF, Yoneda JS, et al. Proteoliposomes in nano-biotechnology. Biophysical reviews. 2012; 4(1):67–81. https://doi.org/10.1007/s12551-011-0065-4 PMID: 28510001; PubMed Central PMCID: PMC5418368.

42. Beglova N, Blacklow SC, Takagi J, Springer TA. Cysteine-rich module structure reveals a fulcrum for integrin rearrangement upon activation. Nature structural biology. 2002; 9(4):282–7. https://doi.org/10.1038/nsb779 PMID: 11896403.

43. Vieira JL, Borges LM, Nascimento MT, Gomes Ade L. Quinine levels in patients with uncomplicated falciparum malaria in the Amazon region of Brazil. The Brazilian journal of infectious diseases: an official publication of the Brazilian Society of Infectious Diseases. 2008; 12(5):353–4. PMID: 19219270.

44. Zhu J, Zhu J, Bougie DW, Aster RH, Springer TA. Structural basis for quinine-dependent antibody binding to platelet integrin alphaIIbbeta3. Blood. 2015; 126(18):2138–45. https://doi.org/10.1182/blood-2015-04-639351 PMID: 26282540; PubMed Central PMCID: PMC4626254.

45. Bougie DW, Peterson J, Rasmussen M, Aster RH. Mechanism of quinine-dependent monoclonal antibody binding to platelet glycoprotein IIb/IIIa. Blood. 2015; 126(18):2146–52. https://doi.org/10.1182/blood-2015-04-643148 PMID: 26353910; PubMed Central PMCID: PMC4626255.

46. Bougie DW, Wilker PR, Aster RH. Patients with quinine-induced immune thrombocytopenia have both "drug-dependent" and "drug-specific" antibodies. Blood. 2006; 108(3):922–7. WOS:000239381000029. https://doi.org/10.1182/blood-2006-01-009803 PMID: 16861345

47. Vorup-Jensen T, Chi LL, Gjelstrup LC, Jensen UB, Jewett CA, Xie C, et al. Binding between the Integrin alpha X beta 2 (CD11c/CD18) and heparin. J Biol Chem. 2007; 282(42):30869–77. https://doi.org/10.1074/jbc.M706114200 WOS:000250136300055. PMID: 17699512

48. Ballut L, Sapay N, Chautard E, Imberty A, Ricard-Blum S. Mapping of heparin/heparan sulfate binding sites on alpha v beta 3 integrin by molecular docking. J Mol Recognit. 2013; 26(2):76–85. https://doi.org/10.1002/jmr.2250 WOS:000313836200004. PMID: 23334915

49. Kreimann M, Brandt S, Krauel K, Block S, Helm CA, Weitschies W, et al. Binding of anti-platelet factor 4/heparin antibodies depends on the thermodynamics of conformational changes in platelet factor 4. Blood. 2014; 124(15):2442–9. https://doi.org/10.1182/blood-2014-03-559518 PMID: 25150299.

50. Erb EM, Engel J. Reconstitution of functional integrin into phospholipid vesicles and planar lipid bilayers. Methods in molecular biology. 2000; 139:71–82. https://doi.org/10.1385/1-59259-063-2:71 PMID: 10840779.

51. Parmar MM, Edwards K, Madden TD. Incorporation of bacterial membrane proteins into liposomes: factors influencing protein reconstitution. Biochimica et biophysica acta. 1999; 1421(1):77–90. PMID: 10561473.

52. Bohm G, Muhr R, Jaenicke R. Quantitative analysis of protein far UV circular dichroism spectra by neural networks. Protein engineering. 1992; 5(3):191–5. PMID: 1409538.

53. Lau TL, Kim C, Ginsberg MH, Ulmer TS. The structure of the integrin alphaIIbbeta3 transmembrane complex explains integrin transmembrane signalling. The EMBO journal. 2009; 28(9):1351–61. https://doi.org/10.1038/emboj.2009.63 PMID: 19279607; PubMed Central PMCID: PMC2683045.

54. UniProt C. UniProt: a hub for protein information. Nucleic acids research. 2015; 43(Database issue): D204–12. https://doi.org/10.1093/nar/gku989 PMID: 25348405; PubMed Central PMCID: PMC4384041.

55. Mark James Abraham TM, Roland Schulz, Szilárd Páll, Jeremy C. Smith, Berk Hess, Erik Lindahla. GROMACS: High performance molecular simulations through multi-level parallelism from laptops to supercomputers. SoftwareX. 2015;Volume 1– 2:19–25. https://doi.org/10.1016/j.softx.2015.06.001.

56. Jorgensen WL, Chandrasekhar J, Madura JD, Impey RW, Klein ML. Comparison of Simple Potential Functions for Simulating Liquid Water. J Chem Phys. 1983; 79(2):926–35. https://doi.org/10.1063/1.445869 WOS:A1983QZ31500046.

57. Lindorff-Larsen K, Piana S, Palmo K, Maragakis P, Klepeis JL, Dror RO, et al. Improved side-chain torsion potentials for the Amber ff99SB protein force field. Proteins. 2010; 78(8):1950–8. https://doi.org/10.1002/prot.22711 PMID: 20408171; PubMed Central PMCID: PMC2970904.

58. Khoury GA, Smadbeck J, Tamamis P, Vandris AC, Kieslich CA, Floudas CA. Forcefield_NCAA: ab initio charge parameters to aid in the discovery and design of therapeutic proteins and peptides with unnatural amino acids and their application to complement inhibitors of the compstatin family. ACS synthetic biology. 2014; 3(12):855–69. https://doi.org/10.1021/sb400168u PMID: 24932669; PubMed Central PMCID: PMC4277759.

59. Kirschner KN, Yongye AB, Tschampel SM, Gonzalez-Outeirino J, Daniels CR, Foley BL, et al. GLYCAM06: a generalizable biomolecular force field. Carbohydrates. Journal of computational chemistry. 2008; 29(4):622–55. https://doi.org/10.1002/jcc.20820 PMID: 17849372; PubMed Central PMCID: PMC4423547.

60. Jambeck JP, Lyubartsev AP. Another Piece of the Membrane Puzzle: Extending Slipids Further. Journal of chemical theory and computation. 2013; 9(1):774–84. https://doi.org/10.1021/ct300777p PMID: 26589070.

61. Essmann U, Perera L, Berkowitz ML, Darden T, Lee H, Pedersen LG. A Smooth Particle Mesh Ewald Method. J Chem Phys. 1995; 103(19):8577–93. https://doi.org/10.1063/1.470117 WOS: A1995TE36400026.

62. Pall S, Hess B. A flexible algorithm for calculating pair interactions on SIMD architectures. Comput Phys Commun. 2013; 184(12):2641–50. https://doi.org/10.1016/j.cpc.2013.06.003 WOS:000328725200003.

63. Hess B, Bekker H, Berendsen HJC, Fraaije JGEM. LINCS: A linear constraint solver for molecular simulations. Journal of computational chemistry. 1997; 18(12):1463–72. https://doi.org/10.1002/(Sici)1096-987x(199709)18:12<1463::Aid-Jcc4>3.0.Co;2-H WOS:A1997XT81100004.

64. Bussi G, Donadio D, Parrinello M. COMP 8-Canonical sampling through velocity rescaling. Abstr Pap Am Chem S. 2007; 234. WOS:000207593906171.

65. Parrinello M, Rahman A. Polymorphic Transitions in Single-Crystals—a New Molecular-Dynamics Method. J Appl Phys. 1981; 52(12):7182–90. https://doi.org/10.1063/1.328693 WOS: A1981MT07800024.

66. Roe DR, Cheatham TE. PTRAJ and CPPTRAJ: Software for Processing and Analysis of Molecular Dynamics Trajectory Data. Journal of chemical theory and computation. 2013; 9(7):3084–95. https://doi.org/10.1021/ct400341p WOS:000321793100024. PMID: 26583988